\documentclass[12pt, draftclsnofoot, onecolumn]{IEEEtran}
\ifCLASSINFOpdf
\else
\fi
\usepackage{graphicx}
\usepackage{setspace}
\usepackage{amsmath}
\usepackage{amssymb} 
\usepackage{bm} 
\usepackage{cite}
\usepackage{float}
\usepackage[usenames,dvipsnames]{pstricks}
\usepackage{epsfig}
\usepackage{pst-grad} 
\usepackage{pst-plot} 
\allowdisplaybreaks
\newcommand{\lb} {\left}
\newcommand{\rb} {\right}
\newcommand{\nn} {\nonumber}

\begin{document}
\title{Secrecy Outage  of Dual-hop Regenerative Multi-relay  System with Relay Selection}
\author{Chinmoy Kundu,~\IEEEmembership{Student Member, IEEE,}
        Sarbani Ghose,~\IEEEmembership{Student Member, IEEE,}
        and Ranjan Bose,~\IEEEmembership{Senior Member, IEEE}
\thanks{Chinmoy Kundu is with the Bharti School of Telecommunication Technology \& Management, 
Indian Institute of Technology Delhi, New Delhi-110016, India, 
e-mail: chinmoy.kundu@dbst.iitd.ac.in.}
\thanks{Sarbani Ghose and Ranjan Bose are with the Department 
of Electrical Engineering, Indian Institute of Technology Delhi, New Delhi-110016, India, e-mail: sarbani.ghose@ee.iitd.ac.in, 
rbose@ee.iitd.ac.in.}}
\maketitle
\thispagestyle{empty}
\pagestyle{empty}

\begin{abstract}
Relay selection is considered to enhance the secrecy of a dual-hop regenerative multi-relay system  
with an eavesdropper. Without assuming perfect decoding at the relays, the secrecy outage probability of a single relay system 
is obtained first. Secrecy outage of optimal, traditional and suboptimal relay selection schemes is then evaluated. 
To reduce the power consumption, partial relay selection schemes based only on either of the source-relay or 
relay-destination instantaneous channel state information (ICSI) are introduced. 
Its secrecy outage is evaluated and compared with the other schemes. Secrecy outage of all the selection schemes are 
obtained in closed-form. An optimal relay selection scheme is proposed using secrecy outage which does not require any ICSI. 
Asymptotic and diversity gain analysis of the secrecy outage is presented when source-relay and relay-destination 
average SNRs are same or different. We observe that the improvement in eavesdropper link quality 
affects the secrecy outage more when required secrecy rate is low as compared to the case  when rate is high.  
We also observe that relay selection improves performance more when number of relays are more.
It is important to note that either of the source-relay or the relay-destination  link quality 
can equally limit the secrecy outage performance even if the other link quality is infinitely good.
\end{abstract}
\section {Introduction}
\label{sec_intro}
Wireless medium is broadcast in nature hence any unintended receiver (eavesdroppers) 
can listen to the signals emanating from a source and can potentially be a threat to secure 
communication \cite{wyner_wiretap, Gamal_On_the_Sec_Cap_Fad_Ch}. 
Upper layer data encryption has been  the traditional technique for secure communication. Recently 
researchers have started studying physical layer techniques for information security extensively 
\cite{poor_infor_theo_sec, McLaughlin_wireless_info_theo_sec, Rodrigues_sec_cap_wire_ch, Csiszar_broad_ch_with_conf}. 

Cooperation between nodes is a popular trend in wireless communication as 
it can provide extended coverage and can increase spatial diversity without increasing the number of antennas 
\cite{ Laneman_Wornell_energy_efficient, Laneman_Wornell_cooperative_diversity, hunter_outage_analysis_of_coded}. 
To overcome the wireless channel impairments and improve the performance of secure communications, 
cooperation is also introduced in physical layer security \cite{Gamal_The_Relay_Eaves_Ch_Coop_Sec, 
Petropulu_Poor_Impr_Wire_Phylay_Sec, Petropulu_On_Coop_Rel_Scheme, bao_multihop_sec}. 
The four-terminal relay-eavesdropper channel is introduced in \cite{Gamal_The_Relay_Eaves_Ch_Coop_Sec}. 
Various cooperation strategies like noise-forwarding, compress-and-forward, and amplify-and-forward (AF) 
are discussed and the corresponding achievable performance bounds are derived. Using AF and or 
decode-and-forward (DF) relays in dual-hop cooperative multi-relay system,  \cite{Petropulu_Poor_Impr_Wire_Phylay_Sec} 
and \cite{Petropulu_On_Coop_Rel_Scheme} optimize the achievable 
secrecy rate or the total transmit power. Cooperative jamming is also introduced by 
\cite{Petropulu_Poor_Impr_Wire_Phylay_Sec, Petropulu_On_Coop_Rel_Scheme}  in which the source 
transmits the encoded signal and relays transmit a weighted jamming signal to confuse the eavesdroppers.  

Relay selection   based on instantaneous channel state information (ICSI), as proposed in 
\cite{Khisti_a_simple_cooperative_diversity}, is a novel technique in wireless communication. 
It can increase the diversity gain in a cooperative system without using multiple antennas 
or distributed space-time codes. 
The problem of inefficient spectral utilization of cooperative relays 
in \cite{ Laneman_Wornell_energy_efficient, Laneman_Wornell_cooperative_diversity, hunter_outage_analysis_of_coded}
due to transmission on orthogonal channels, can be overcome by relay selection. 
Relay selection can achieve diversity-multiplexing trade-off. 
In a dual-hop DF multi-relay cooperative system, outage probability and average channel capacity are derived for the 
best relay selection technique in \cite{ikki_relsel}. The best relay in \cite{ikki_relsel} is the relay which 
produces highest signal-to-noise ratio (SNR) at the destination.  In \cite{krikidis_maxmin_relsel}, 
max-min relay selection is considered in an 
interference-limited AF cooperative scenario. Relay selection is also studied recently  in cognitive radio 
\cite{zou_cognitive_relsel} and free space optical communication systems \cite{schober_relsel_FSO}. 

Optimal relay selection requires monitoring ICSI globally \cite{Khisti_a_simple_cooperative_diversity, ikki_relsel}. 
Monitoring partial ICSI among the nodes locally as opposed to globally 
can reduce the complexity and power consumption, hence can prolong the lifetime of the network \cite{krikidis_partial_relay_selection, 
rui_capacity_analysis_df, adve_selection_cooperation}. 
In \cite{krikidis_partial_relay_selection}, partial relay selection method is introduced to select a relay among multiple relays 
in a two-hop AF system using only source-relay ICSI. In \cite{rui_capacity_analysis_df}, partial relay selection is 
extended for DF cooperative networks to find exact expressions of capacity performance. 
In \cite{adve_selection_cooperation}, along with optimal relay selection method, 
a partial relay selection method is also investigated in bandwidth-limited wireless systems with multiple sources. 
Contrary to source-relay ICSI based selection \cite{Khisti_a_simple_cooperative_diversity, ikki_relsel, rui_capacity_analysis_df}, 
in \cite{adve_selection_cooperation} the destination selects the relay with the best ICSI of the relay-destination link. 
Approximate outage probability in high-SNR is obtained and it is shown that it can outperform distributed space-time codes. 
All the literature for relay selection discussed till now deals with relay selection in general communication scenario 
without any eavesdropper.

Recently relay selection in cooperative physical layer security has got considerable attention in order to improve 
security against eavesdropping attack \cite{krikidis_iet_opport_rel_sel, 
krikidis_twc_Rel_Sel_Jam, Hui_Secure_Relay_Jammer_Selection, Zou_Wang_Shen_optimal_relay_sel, 
Alotaibi_Relay_Selection_MultiDestination, Bao_Relay_Selection_Schemes_Dual_Hop_Security, Poor_Security_Enhancement_Cooperative, 
Fan_Karagiannidis_Secure_Multiuser_Communications, 
jindal_kundu_secrecy_paper_vtc_2014, jindal_kundu_secrecy_af_coml, DBLP_Zou_Outage_Cognitive_Radio, Qahtani_relsel}.
Based on the available knowledge of the ICSI or the statistical channel state information (SCSI) of all 
the links, the problem of relay selection 
is mostly solved for the following three cases. For the case i), when all the ICSI is known optimal relay selection 
is implemented to find the relay for which secrecy capacity is maximum.  
In case ii), when all but relay-eavesdropper ICSI is known, traditional relay selection is implemented.  
Traditional relay selection does not measure secrecy capacity instead it measures main channel capacity only. 
In case iii), when all but relay-eavesdropper ICSI is known and relay-eavesdropper SCSI is also known, suboptimal relay selection 
is performed. Relay selection for the cases when no ICSI is known or ICSI of either one of the source-relay or relay-destination is known, 
are not discussed.

For the dual-hop multiple DF relay system in \cite{krikidis_iet_opport_rel_sel}, 
secrecy outage probability is derived for the selection schemes from i) to iii). Assuming high SNR where all the 
relay nodes successfully decode the source transmission, \cite{krikidis_iet_opport_rel_sel} simplifies 
the analysis. To improve the system in \cite{krikidis_iet_opport_rel_sel}  a jammer is selected along with 
the relay in \cite{krikidis_twc_Rel_Sel_Jam} and secrecy outage probability for the same is obtained. 
By taking multiple antenna destination in a dual-hop cooperative system with 
multiple eavesdroppers, relay and jammer selection methods are also studied 
in \cite{Hui_Secure_Relay_Jammer_Selection} to minimize secrecy outage probability. 
In \cite{Zou_Wang_Shen_optimal_relay_sel}, closed-form intercept probability expressions for relay selection 
schemes of  i) and ii)  are derived using both the DF and AF relays in dual-hop multi-relay system. 
It finds non-zero secrecy capacity which is more tractable than the secrecy outage probability. 
In \cite{Alotaibi_Relay_Selection_MultiDestination}, outage probability of the optimal relay selection scheme 
is obtained for dual-destination case with a single eavesdropper having DF relays in dual-hop scenario.
Instead of single eavesdropper as in \cite{krikidis_iet_opport_rel_sel, krikidis_twc_Rel_Sel_Jam, 
Zou_Wang_Shen_optimal_relay_sel, Alotaibi_Relay_Selection_MultiDestination}, 
the effects of relay selection schemes of i) and ii) are is studied in \cite{Bao_Relay_Selection_Schemes_Dual_Hop_Security}. 
  Authors in \cite{Bao_Relay_Selection_Schemes_Dual_Hop_Security} obtain probability of non-zero achievable 
secrecy rate, secrecy outage probability and achievable secrecy rate for multiple eavesdroppers using 
dual hop multiple DF relay system. 
A two-stage relay and destination selection procedure is investigated in \cite{Poor_Security_Enhancement_Cooperative} 
for  cooperative single carrier systems with multiple eavesdroppers and destinations having multiple DF relays. 
Several security metrics like the secrecy outage probability, probability of nonzero secrecy rate, and ergodic secrecy 
rate is obtained in frequency selective fading channel. 

Having multiple eavesdroppers and destinations, \cite{Fan_Karagiannidis_Secure_Multiuser_Communications} 
presents three criteria to select the best relay and user pair which maximizes the secrecy outage probability. 
Multiple AF relays are considered in dual-hop cooperative scenario instead of DF relays.
Secrecy outage probability of dual-hop AF relay system with single eavesdropper is investigated in \cite{jindal_kundu_secrecy_af_coml, 
jindal_kundu_secrecy_paper_vtc_2014}. No DF relay is considered here.  In \cite{jindal_kundu_secrecy_af_coml}, relay selection 
is considered when ICSI of the eavesdropper is not available. In \cite{jindal_kundu_secrecy_paper_vtc_2014}, an optimal 
relay selection method based on secrecy outage probability is proposed which does not require any ICSI measurement. 
In cognitive radio network, cognitive user scheduling can be found to improve the secrecy and diversity in multiple eavesdropper and 
multiple cognitive user system  \cite{DBLP_Zou_Outage_Cognitive_Radio}. Secrecy outage probability and corresponding 
diversity gain is obtained for various scheduling schemes depending on the available ICSIs.

Whenever relay selection  problem is considered for secrecy in cooperative DF relaying 
\cite{krikidis_iet_opport_rel_sel, krikidis_twc_Rel_Sel_Jam, 
Bao_Relay_Selection_Schemes_Dual_Hop_Security, Alotaibi_Relay_Selection_MultiDestination, 
Poor_Security_Enhancement_Cooperative, Hui_Secure_Relay_Jammer_Selection}, perfect decoding 
is assumed at each relay in the high SNR scenario. 
By doing so,  the effects of the quality of the first hop link is neglected which actually can affect the 
rate of the particular branch to the destination and the secrecy rate. 
Actually later in the numerical section  we have shown that both the source-relay and relay-destination 
link quality can equally affect the performance. 
Hence in this paper we concentrate on the cooperative multiple DF relay scenario 
without considering high SNR scenario and constraints on the relays that they correctly decode the messages. 
Though the high SNR assumption is not considered in \cite{Zou_Wang_Shen_optimal_relay_sel}, authors 
do not obtain secrecy outage probability instead  they obtain non-zero secrecy capacity. 
Non-zero secrecy capacity is more tractable than the secrecy outage probability which we derive in this work. 
Deviating from the fact of perfect decoding at all the relays, \cite{Qahtani_relsel} considers only a set 
of relays can successfully decode the message among all the relays. Relays can decode the message 
only if the SNR at them meet a predetermined threshold. In this paper 
relay selection is performed only on the basis of relay-destination link quality of the successfully decoded relays.
This makes the system significantly different from our system as no threshold based decoding is considered 
in our paper. Not only that we consider both source-relay and relay-destination link qualities for relay selection 
along with other relay selection methods.

Optimal relay selection requires global ICSI which is complex to obtain and power consuming. 
By acquiring partial ICSI, power consumption can be reduced and lifetime of the network can be increased 
\cite{krikidis_partial_relay_selection, rui_capacity_analysis_df, adve_selection_cooperation}. 
Even global ICSI may not be available at all the time at a central unit and relay selection 
might require to be performed in a distributed manner. 
Hence partial relay selection schemes based on either of the source-relay or relay-destination ICSI is introduced 
in this paper in secrecy setup to enhance the secrecy. Partial relay selection is not present in 
\cite{krikidis_iet_opport_rel_sel, krikidis_twc_Rel_Sel_Jam, Bao_Relay_Selection_Schemes_Dual_Hop_Security, 
Zou_Wang_Shen_optimal_relay_sel, Alotaibi_Relay_Selection_MultiDestination, Poor_Security_Enhancement_Cooperative, 
Fan_Karagiannidis_Secure_Multiuser_Communications, Hui_Secure_Relay_Jammer_Selection}. 
Apart from DF relays, secrecy enhancement by relay selection in dual-hop cooperative system with 
AF relays is discussed in \cite{Fan_Karagiannidis_Secure_Multiuser_Communications, 
jindal_kundu_secrecy_af_coml, jindal_kundu_secrecy_paper_vtc_2014}. These work are different from our work as we have 
considered DF relaying in this paper. Our work is also significantly different from 
the work in \cite{DBLP_Zou_Outage_Cognitive_Radio} where authors do not consider cooperative relaying 
in cognitive radio network whereas we consider cooperative relaying in non-cognitive system.

The motivations of this work can be outlined as follows. Relay selection can improve the diversity performance 
without using complex multiple antenna system. Not only that it can achieve diversity-multiplexing trade-off. 
Hence relay selection is considered to improve the secrecy in a cooperative dual-hop system with a single eavesdropper.  
Whenever relay selection is considered in dual-hop DF relay system with secrecy, it is assumed that relays can correctly 
decode the message due to high SNR at the first hop. Which may not be true for all the SNR regimes. 
Motivated by this we have assumed that relays can make incorrect decisions. In this case rate through the 
particular branch is limited by the minimum quality of the individual hop among the dual-hop link. 
Secrecy outage probability is obtained using this assumption for various relay selection schemes in closed form. 
Power consumption is an important issue in a energy constrained network like sensor network. Optimal relay selection 
requires global  ICSI measurement. Even traditional relay selection requires source-relay and relay-destination 
ICSI. ICSI measurement is complex and power consuming and decreases the lifetime of any network. 
To increase the lifetime of the network in a energy constrained network, partial relay selection 
is introduced. It requires less channel measurements can reduce power consumption but even improves secrecy 
due to selection. Power consumption can be reduced further if selection process does not requires any 
ICSI measurement. Secrecy outage probability is such a metric and an optimal relay selection is proposed 
minimizing this.

In this paper, relay selection is considered to enhance the physical layer security in a dual-hop multi-relay system  
with one source, one destination, multiple DF relays and an eavesdropper. 
We obtain secrecy outage probability of a single relay system first. Secrecy outage probability of relay selection schemes for all 
the three cases (i-iii) discussed earlier are derived then. 
Secrecy outage probabilities of partial relay selection schemes are derived and compared with the other selection 
schemes (i-iii). All the secrecy outage probabilities are presented in closed-form. 
An optimal relay selection scheme is proposed with the help of  secrecy outage probability of the single relay system which 
only requires SCSI of the links. As no complex ICSI measurement is required it can reduce power consumption and 
complexity even more than the partial relay selection.  
Asymptotic analysis is presented and diversity order is determined for the secrecy outage probability of the single relay system 
and multiple relay system with relay selection. Asymptotic analysis is provided when average SNRs of 
source-relay and relay-destination links are equal or unequal. When equal, we call it as balanced case and when unequal, we call it as 
unbalanced case. It is observed that improvement in eavesdropper channel quality 
affects the secrecy outage probability more when required threshold rate is low as compared to the case 
when it is high. We also observe that relay selection improves the performance more when number of relays are more.
It is interesting to find that either of the source-relay or relay-destination link quality can equally 
limit the secrecy outage probability even if the other link average SNR is infinitely good.

The main contribution of the paper can be summarized as
\begin{itemize}
 \item Without assuming that the DF relays can always decode the message correctly, we evaluate secrecy outage probability of 
various relay selection schemes in closed form for the dual-hop cooperative DF relay system. 
We are able to show that both the source-relay or relay-destination 
link quality can equally affect the secrecy outage performance.

\item Depending on the availability of the ICSI and to reduce the power consumption or increase the lifetime of a 
network, we introduce partial relay selection  in a simple dual-hop cooperative DF relay system to enhance the secrecy 
outage probability. 

\item We provide asymptotic analysis for the secrecy outage probability of the single relay system or the 
multi-relay system with relay selection for the balanced and unbalanced cases. From the asymptotic analysis we 
show that the relay selection improves diversity gain of the system.

\end{itemize}

The rest of the paper is organized as follows. Section \ref{sec_proposed} describes the system model. 
Secrecy outage probability expressions are derived for single relay system 
in section \ref{sec_out_prob}. Secrecy outage probability of different relay selection schemes are presented 
in section \ref{out_prob_rel_sel}. Asymptotic analysis are provided in section \ref{sec_asymp}. 
Simulation results are discussed in section \ref{sec_results}. Finally, the paper is concluded in section \ref{sec_conclusion}.  

\textit{Notation:} $\mathcal{E} \left(x\right)$ defines exponential distribution with  parameter $x$, 
 $\mathbb{P}[\cdot]$ is the probability of an event, $\mathbb{E}_X[\cdot]$ is the expectation of its argument 
over random variable (RV) $X$. $\max\{\cdot\}$ and $\min\{\cdot\}$ denote the maximum and minimum of its arguments respectively 
and $(x)^+\triangleq \max(0,x)$. Generally $F_X(\cdot)$, in capital letter, denotes the cumulative distribution 
function (CDF) of a RV $X$. $f_X(\cdot)$ , in small letter, denotes the corresponding probability density function (PDF).

\section {System Model}
\label{sec_proposed}
\begin{figure} 
\centering
\includegraphics[width=0.275\textwidth] {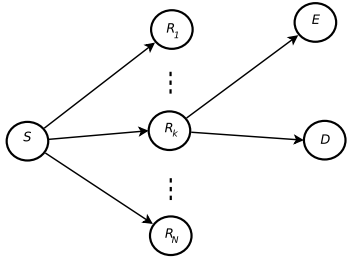}
\vspace*{-0.1cm}
\caption{Dual-hop multi-relay regenerative system where one of the relay is selected to forward the source information.}
\label{FIG_1}
\vspace{-0.4cm}
\end{figure}

The system model, as depicted in the Fig.\ref{FIG_1}, consists of one source ($S$), one destination ($D$), and one 
passive eavesdropper ($E$), and $N$ number of regenerative or DF relays working in a dual-hop mode. 
High SNR assumption at the relays as in \cite{krikidis_iet_opport_rel_sel, krikidis_twc_Rel_Sel_Jam} 
 or all DF relays can perfectly decode the source 
information  as  in \cite{Bao_Relay_Selection_Schemes_Dual_Hop_Security, 
Alotaibi_Relay_Selection_MultiDestination, Poor_Security_Enhancement_Cooperative, Hui_Secure_Relay_Jammer_Selection} 
has not been made. 
To focus our study on the cooperative slot, the direct links between $S$-$D$ and $S$-$E$ are not considered 
assuming direct links are in deep shadow fading or the nodes may be far apart. It is worth noting that this assumption 
is very well known not only in cooperative communication but also in cooperative secrecy communication scenarios 
\cite{Bletsas_Outage_analysis_cooperative_communication, Eddaghel_Outage_amplifyforward_cooperative, 
Poor_Secure_wireless_communications, Poor_Secrecy_capacity_class_of_orthogonal}.
This assumption is considered recently in the dual-hop cooperative secrecy setup without multiple eavesdropper 
in \cite{krikidis_iet_opport_rel_sel, krikidis_twc_Rel_Sel_Jam, 
Zou_Wang_Shen_optimal_relay_sel, 
Alotaibi_Relay_Selection_MultiDestination} 
and with multiple eavesdroppers in 
\cite{Hui_Secure_Relay_Jammer_Selection, Bao_Relay_Selection_Schemes_Dual_Hop_Security, 
Poor_Security_Enhancement_Cooperative, Fan_Karagiannidis_Secure_Multiuser_Communications}. 
This assumption is also reasonable for the cooperative system with secure broadcast 
phase \cite{Poor_Secrecy_capacity_class_of_orthogonal} or the system where source node communicates 
with the relay node via a local connection \cite{Ozgur_Tse_Hierarchical_Cooperation_Achieves}.

In the first time slot the messages from the $S$ is decoded at the relay. In the second time slot, 
a relay is selected among $N$ available relays. It re-encodes the message and forwards it to the destination. 
The  links between various nodes are modeled as mutually independent flat Rayleigh fading but not identical. 
The SNR between any two arbitrary nodes $x$ and $y$, denoted as $\gamma_{xy}$, is given by 
\begin{align}
\label{eq_1} 
\gamma_{xy} = \frac{P_{x}h_{xy}^2 }{N_{0_{y}}},
\end{align}
where $P_{x}$ is the transmit power at node $x$, $N_{0_{y}}$ is the noise variance of the additive white Gaussian noise at $y$. 
As $h_{xy}$ is Rayleigh distributed, $\gamma_{xy}$ is exponential distributed with mean $1/\beta_{xy}$ \cite{book_proakis}, 
denoted as $\gamma_{xy} \sim\mathcal{E} \left(\beta_{xy}\right)$, where $\beta_{xy}$ is the parameter of the exponentially distribution. 
The PDF of the exponential distribution with parameter $\beta_{xy}$, is
\begin{align}
\label{eq_1_A1} 
f_X(z) =\beta_{xy}e^{-z\beta_{xy}}, 
\end{align}
and corresponding CDF is 
\begin{align}
\label{eq_1_A1} 
F_X(z) =1-e^{-z\beta_{xy}}. 
\end{align} 
The achievable secrecy rate is defined as \cite{wyner_wiretap, Rodrigues_sec_cap_wire_ch, Gamal_The_Relay_Eaves_Ch_Coop_Sec, 
Zou_Wang_Shen_optimal_relay_sel}
\begin{align}
\label{eq_2}
C_S\triangleq{\frac{1}{2}\lb[\log_2\lb(\frac{1+\gamma_{M}}{1+\gamma_{E}}\rb)\rb]}^+,
\end{align} 
where $\gamma_{M}$ and $\gamma_{E}$ are the SNRs at $D$ and $E$ respectively. 
$\gamma_{M}$ and $\gamma_{E}$ are termed as the main channel and the eavesdropper channel SNRs respectively. 
The term $1/2$ denotes that two time slots are required to complete the transmission process. 
Secrecy outage probability which is defined as the probability that the instantaneous secrecy rate is less than 
a desired threshold secrecy rate  for the system is \cite{Rodrigues_sec_cap_wire_ch}
\begin{align} 
\label{eq_3}
P_o\lb(R_s\rb)=\mathbb{P}\left[C_S < R_s \right]= \mathbb{P}\left[\frac{1+\gamma_{M}}{1+\gamma_{E}}<\rho\right],
\end{align} 
where, $R_s>0$ is the desired threshold secrecy rate of the system and $\rho =2^{2R_s}$.  As $\rho$ is a direct mapping of 
desired threshold secrecy rate $R_s$, we use both the terms as threshold secrecy rate throughout the paper interchangeably.                                   

To distinguish $S$-$R_k$, $R_k$-$D$ and $R_k$-$E$ links, for all $k=1, \cdots, N$, 
we replace the subscripts $xy$ with $sk$, $kd$ and $ke$ respectively.   
Throughout the paper parameter of exponential distribution 
for $S$-$R_k$, $R_k$-$D$ and $R_k$-$E$  links,  for all $k=1, \cdots, N$, are assumed to be 
$\beta_{sk}$, $\beta_{kd}$ and $\alpha_{ke}$ respectively.

It should be noted that with an assumption of DF relays 
can have decoding errors, a more general version of the relay selection problem 
with multiple eavesdroppers can be considered 
as in \cite{DBLP_Zou_Outage_Cognitive_Radio, Bao_Relay_Selection_Schemes_Dual_Hop_Security, 
Fan_Karagiannidis_Secure_Multiuser_Communications, Poor_Security_Enhancement_Cooperative, 
Hui_Secure_Relay_Jammer_Selection} with multiple destinations. 
The direct links from the source to the destination and eavesdropper can also be considered to study 
whether new relay selection criteria can be proposed. 
In case of direct links, the performance of various diversity combining techniques at the eavesdropper 
and destination can be investigated further.

\section {Secrecy Outage of Dual-hop single relay System}  
\label{sec_out_prob}
Considering high SNR scenario, \cite{krikidis_iet_opport_rel_sel, krikidis_twc_Rel_Sel_Jam} assumes that the DF relay can correctly 
decode the information. So  in (\ref{eq_2}), for the $k^{th}$ relay, \cite{krikidis_iet_opport_rel_sel, krikidis_twc_Rel_Sel_Jam} 
consider $\gamma_M= \gamma_{kd}$. 
Relay selection  problems for the cooperative DF relaying in \cite{Bao_Relay_Selection_Schemes_Dual_Hop_Security, 
Alotaibi_Relay_Selection_MultiDestination, Poor_Security_Enhancement_Cooperative, Hui_Secure_Relay_Jammer_Selection}, also consider 
perfect decoding at each relay. 
This neglects the fact that quality of the first hop link can also affect the rate of the 
particular branch and eventually secrecy rate. The assumption of perfect decoding at each relay 
may not be appropriate for all the SNR regimes. Instead, for the  DF relay system, 
the capacity of dual-hop system is limited by the minimum of the individual hop capacities. This corresponds to the
minimum of the individual hop SNRs. This assumption takes care of the fact that DF relays can have decoding errors. 
Motivated by this we take the secrecy rate of a single DF relay system, i.e. for the $k^{th}$ relay, following 
\cite{Zou_Wang_Shen_optimal_relay_sel} as

\begin{equation}
\label{eq_5}
C_S^k  = {\frac{1}{2}\lb[\log_2\lb(\frac{1+\gamma_k}{1+\gamma_{ke}}\rb)\rb]}^+,
\end{equation}
where $\gamma_k$ is the minimum of the $S$-$R_k$ and $R_k$-$D$ link SNRs, is given  by
\begin{equation}
\label{eq_4}
\gamma_k=\min\lb(\gamma_{sk},\gamma_{kd}\rb). 
\end{equation}
In \cite{Zou_Wang_Shen_optimal_relay_sel}, $\gamma_M$ in (\ref{eq_2}) is replaced by $\gamma_k$ and $\gamma_E$ by $\gamma_{ke}$.

Using (\ref{eq_5}) in (\ref{eq_3}), we evaluate the secrecy outage probability for the $k^{th}$ relay as
\begin{align}
\label{eq_6}
P_o^k(R_s)&=\mathbb{P}\lb[C_S^k<R_s\rb]=\mathbb{P}\left[\gamma_k < \rho \left(1+\gamma_{ke}\right)-1\right] \nn \\
&=\mathbb{P}\lb[\min\lb(\gamma_{sk},\gamma_{kd}\rb)<\lambda\rb] \nn\\
&=1-\mathbb{P}\lb[\min\lb(\gamma_{sk},\gamma_{kd}\rb)\ge\lambda\rb] \nn \\
&=1- \mathbb{P}\lb[\gamma_{sk}\ge\lambda\rb] \mathbb{P}\lb[\gamma_{kd}\ge\lambda\rb] \nn \\
&=1- \lb(1-\mathbb{P}\lb[\gamma_{sk}<\lambda\rb] \rb)\lb(1-\mathbb{P}\lb[\gamma_{kd}<\lambda\rb] \rb)\nn \\
&= \mathbb{E}_{\gamma_{ke}}\lb[  F_{\gamma_{sk}} \lb(\lambda\rb) + F_{\gamma_{kd}}\lb(\lambda\rb) - 
   F_{\gamma_{sk}}\lb(\lambda\rb)F_{\gamma_{kd}}\lb(\lambda\rb) \rb]\nn \\
&=\int_0^\infty \lb(1-e^{-\lb(\beta_{sk}+\beta_{kd}\right)\lambda}\rb)\alpha_{ke}e^{-\alpha_{ke}\gamma_{ke}}d\gamma_{ke} \nn\\
&=1-\frac{\alpha_{ke} e^{-\lb(\beta_{sk}+\beta_{kd}\rb)\lb(\rho-1\rb)}}{\rho \lb(\beta_{sk}+\beta_{kd}\right)+\alpha_{ke}},
\end{align}
where $\lambda=\rho \left(1+\gamma_{ke}\right)-1$, $F_{\gamma_{sk}}(\cdot)$ and $F_{\gamma_{kd}}(\cdot)$ 
are exponential CDFs with parameters $\beta_{sk}$ and $\beta_{kd}$ respectively. 
\section {Secrecy Outage of Relay Selection Schemes} 
\label{out_prob_rel_sel} 
This section evaluates the secrecy outage probability of three relay selection schemes (i-iii) 
as discussed in section \ref{sec_intro}. To reduce the system complexity and increase network lifetime by 
decreasing power consumption, two partial relay selection schemes are introduced in the context of 
cooperative physical layer security. The secrecy outage probabilities are obtained for the same. 
All the secrecy outage probabilities obtained in this section are without assuming high 
SNR scenario as opposed to \cite{krikidis_iet_opport_rel_sel, krikidis_twc_Rel_Sel_Jam}. 
In a power and computational complexity constrained network where no ICSI is available, an optimal relay selection method 
is also proposed. This optimal relay selection method is based on secrecy outage probability of dual-hop single 
relay system which is derived in the previous section.
\subsection{Optimal Selection: ICSI of all links are known ($OS$)}
\label{subsec_optimal}
In the optimal relay selection 
\cite{krikidis_iet_opport_rel_sel, krikidis_twc_Rel_Sel_Jam, Zou_Wang_Shen_optimal_relay_sel}, 
the relay is selected for which the achievable secrecy rate of the system becomes maximum. 
In this case, secrecy outage probability can be obtained by finding the probability for which the maximum 
achievable secrecy rate of the system is less than the required threshold. 
The probability of the maximum of some independent RVs is less than some quantity 
is the probability that all the RVs are less than that quantity. 
This can be evaluated by multiplying the CDFs of the corresponding RVs. Here CDFs are basically the 
secrecy outage probabilities of individual single relay systems. Hence 
the secrecy outage probability  of the optimal relay selection can be obtained as 
\begin{align}
\label{eq_7}
P_o^{OS}(R_s)&=\mathbb P\left[\max_{k \in [1,N]} \{C_S^k\}< R_s\right]\nn\\
&=\prod_{k=1}^N \mathbb P\left[C_S^k< R_s\rb]   =\prod_{k=1}^N P_o^k(R_s).
\end{align}
We can check that (\ref{eq_7}) is simply the multiplication of secrecy outage probabilities 
of individual single relay system derived in (\ref{eq_6}). 
This relay selection method requires ICSI of all the links i.e. $S$-$R_k$, $R_k$-$D$ and $R_k$-$E$ for all $k=1,\cdots, N$, 
at a central unit to find out the relay with maximum secrecy rate. How the ICSI is 
acquired and relay selection is implemented is beyond the scope of our work and 
can be better understood  from \cite{adam_adaptive_relay_selection , Khisti_a_simple_cooperative_diversity}.
\subsection{Traditional Selection: $R$-$E$ link ICSI is unknown ($TS$)}
\label{subsec_traditional}
Implementing optimal relay selection method in the section \ref{subsec_optimal} requires ICSI  of all the  links beforehand. 
Obtaining ICSI of all the links can be difficult as eavesdropper might be an external entity to the system 
\cite{poor_interference_alignment}. Hence 
traditional relay selection as termed in \cite{Zou_Wang_Shen_optimal_relay_sel} can be an alternative relay selection rule. 
In traditional relay selection the relay is selected for which the main channel secrecy rate becomes maximum. 
In this relay selection method all but eavesdropper link ICSI is required at a central unit. 
The secrecy outage probability can be evaluated by finding conditional secrecy outage probability when a particular relay 
let us say $R_k$ is selected and summing over all such possibilities. The particular relay $R_k$ is selected only if 
the main channel secrecy rate of that particular branch 
or alternatively the instantaneous SNR, $\gamma_k$, is maximum among all other main branch instantaneous SNRs. 
Secrecy outage probability  can be evaluated following the law of total probability as
\begin{align}
\label{eq_12}
&P_o^{TS}(R_s)=\sum\limits_{k=1}^N \mathbb{P}\lb[\text{Relay}=R_k\rb]\mathbb{P}\lb[C_S^k<R_s\rb]  \nn\\
&=\sum\limits_{k=1}^N \mathbb{P}\lb[\gamma_{k}>\gamma_M^{-}\rb]\mathbb{P}\lb[\gamma_{k}<\lambda\rb] \nn\\
&=\sum\limits_{k=1}^N \int_0^\infty \int_0^\lambda \int_y^\lambda 
f_{\gamma_{k}}(x)f_{\gamma_M^{-}}(y)f_{\gamma_{ke}}\lb(z\rb)dx dy dz,
\end{align}
where 
\begin{align}
\label{eq_8}
\gamma_M^{-}&=\max_{\substack{i=1,\dots,N\\i\ne k}}\{ \gamma_i\}, ~~ \text{and} ~~ \gamma_M= \max_{\substack{i=1,\dots,N}}\{\gamma_i\}. 
\end{align}
In (\ref{eq_8}), $\gamma_i$, for all $i$, are defined in (\ref{eq_4}).
The derivation of (\ref{eq_12}) requires the distribution of $\gamma_M^{-}$, $\gamma_k$ and $\gamma_{ke}$. Each link in the system 
 undergoes independent Rayleigh fading so distribution of $\gamma_{ke}$ is exponential. 
The distribution of $\gamma_k$ in (\ref{eq_4}) is exponential with parameter $\beta_k=\beta_{sk}+\beta_{kd}$ 
\cite{book_papoulis}, i.e. $\gamma_k\sim\mathcal{E}(\beta_k)$. The distribution of $\gamma_M^{-}$ can be derived 
from the distribution of $\gamma_M$. 
The distributions of $\gamma_M$  is derived first. The CDF of a RV, which is maximum within some independent RVs, can be 
written as the product of CDFs of the individual RVs \cite{book_papoulis}. Hence $F_{\gamma_M}(x)$ can be written as
\begin{align}
\label{eq_9}
&F_{\gamma_M}(x)=\prod_{m=1}^{N}F_{\gamma_m}(x)=\prod_{m=1}^{N} \lb(1-e^{-\beta_mx}\rb) \nn \\
&=1-\sum_{i_1=1}^{N}e^{-x\beta_{i_1}}
+\sum_{i_1=1}^{N-1}\sum_{i_2=i_1+1}^{N}e^{-x\lb(\beta_{i_1}+\beta_{i_2}\rb)} -\dots+ \nn \\
&(-1)^N \sum_{i_1=1}^{N-(N-1)}\sum_{i_2=i_1+1}^{N-(N-2)}\cdots\sum_{i_N=i_{N-1}+1}^N
e^{-x\lb(\beta_{i_1}+\beta_{i_2}+\dots+\beta_{i_N}\rb)} \nn \\
&=1+\sum_{m=1}^{N}(-1)^m \mathbb {\sum}_m e^{-x\beta_m^{\prime}},
\end{align}
where 
\begin{align}
\label{eq_9_1}
\mathbb {\sum}_m=\sum_{i_1=1}^{N-(m-1)}\sum_{i_2=i_1+1}^{N-(m-2)}\cdots\sum_{i_{m-1}=i_{m-2}+1}^{N-1} \sum_{i_m=i_{m-1}+1}^N,
\end{align}
and $\beta_m^{\prime}=\sum_{l=1}^m\beta_{i_l}$. 
Similarly we can show that 
\begin{align}
\label{eq_15}
F_{\gamma_M^{-}}(x)=1+\sum\limits_{m=1}^{N-1}{\lb(-1\rb)^{m}} \mathbb {\sum}_m^{\prime}e^{-x\beta_m^{\prime}},
\end{align}
where   
\begin{align}
\mathbb {\sum}_m^{\prime}=\sum_{\substack{i_1=1\\i_1\ne k}}^{N-(m-1)}
\sum_{\substack{i_2=i_1+1\\i_2\ne k}}^{N-(m-2)}\cdots \sum_{\substack{i_{m-1}=i_{m-2}+1\\i_{m-1}\ne k}}^{N-1} 
\sum_{\substack{i_m=i_{m-1}+1\\i_m\ne k}}^N.
\end{align}
The PDF of $\gamma_M^{-}$ is found by differentiating $F_{\gamma_M^{-}}(x)$ as
\begin{align}
\label{eq_14}
 f_{\gamma_{M}^-}(x)=-\sum\limits_{m=1}^{N-1}\lb(-1\rb)^{m}\mathbb {\sum}_m^{\prime} \beta_m^{\prime}e^{-x\beta_m^{\prime}}.
\end{align} 
To obtain $P_o^{TS}(R_s)$ in closed-form, we evaluate (\ref{eq_12}) using distributions of $\gamma_{k}$, $\gamma_{ke}$ and 
$\gamma_{M}^-$ from (\ref{eq_14}).
$P_o^{TS}(R_s)$ is finally written in (\ref{eq_16}).   
\subsection {Suboptimal Selection: ICSI of S-R and R-D link and Average R-E link is known ($SS$-$RE$)}
\label{subsec_subopt_re} 
When ICSI of the $R_k$-$E$ for all $k=1, \cdots, N$, links are unavailable, average channel knowledge 
$1/\alpha_{ke}$ can be exploited in traditional relay selection 
rule of section \ref{subsec_traditional}  to improve the performance of it 
\cite{krikidis_iet_opport_rel_sel, krikidis_twc_Rel_Sel_Jam}. This is one of the suboptimal relay selection 
method discussed in this paper. The secrecy outage probability can be evaluated using law of total probability 
as in section \ref{subsec_traditional} as
\begin{align}
\label{eq_17}
&P_{o}^{SS-RE}=
\sum\limits_{k=1}^N \mathbb{P}\lb[\text{Relay}=R_k\rb]\mathbb{P}\lb[C_S^k<R_s\rb]  \nn\\
&=\sum_{k=1}^N \mathbb{P}\lb[\frac{\gamma_k}{1/\alpha_{ke}}>\gamma_M^{-}\rb]\mathbb{P}\lb[\gamma_{k}<\lambda\rb]\nn \\
&=\sum_{k=1}^N\int_0^\infty\int_0^{\lambda\alpha_{ke}}\int_{\frac{y}{\alpha_{ke}}}^\lambda f_{\gamma_k}(x)f_{\gamma_M^{-}}(y)
f_{\gamma_{ke}}(z)dxdydz. 
\end{align}
Following (\ref{eq_8}) in this section, $\gamma_M^{-}$ is defined  as 
\begin{align}
\gamma_M^{-}=\max_{\substack{i=1,\dots,N\\i\ne k}}\lb\{\frac{\gamma_i}{1/\alpha_{ie}}\rb\}
=\max_{\substack{i=1,\dots,N\\i\ne k}}\lb\{\gamma_i\alpha_{ie}\rb\}.
\end{align}
To find the PDF of $\gamma_M^{-}$ we need to know the CDF of $\gamma_i\alpha_{ie}$, which can be found easily as
\begin{align}
\label{eq_19}
\mathbb{P}\lb[{\gamma_i\alpha_{ie}}\le x\rb]
=F_{\gamma_i}\lb(\frac{x}{\alpha_{ie}}\rb).
\end{align} 
Hence we obtain the CDF of $\gamma_M^{-}$ as
\begin{align}
  \label{eq_20}
  \gamma_M^{-}=\prod_{\substack{i=1\\
i\ne k}}^NF_{\gamma_i}\lb(\frac{x}{\alpha_{ie}}\rb).
\end{align}
Equation (\ref{eq_20}) can be written in similar form of (\ref{eq_15}) as a simple summation of exponential functions 
with $\beta_m^{\prime}=\sum_{l=1}^m\beta_{i_l}/\alpha_{le}$. 
Now the solution of (\ref{eq_17}) can  be presented with $\beta_m^{\prime}=\sum_{l=1}^m\beta_{i_l}/\alpha_{le}$ in (\ref{eq_22}).
\subsection {Suboptimal Relay selection: Only R-D link ICSI is known ($SS$-$RD$)}
\label{subsec_subopt_rd}
Global knowledge of instantaneous channel information is power consuming and complex. 
In a resource constrained wireless network like sensor networks, network lifetime can be 
increased and complexity can also be reduced by locally measuring the instantaneous channel 
\cite{krikidis_partial_relay_selection, rui_capacity_analysis_df,adve_selection_cooperation}. 
Hence to reduce the complexity and increase the network lifetime we introduce this partial 
relay selection method in secure communication scenario. 
Only the relay-destination link quality is used to select the best relay in this section. 
This relay selection method is also suboptimal. In a general communication 
scenario without any eavesdropper, relay-destination ICSI based partial relay selection methods 
can be found in \cite{adve_selection_cooperation}, but not in secure communication setup. 
Sometimes it can happen that ICSI of only $R_k$-$D$ links, for all $k=1, \cdots, N$, is available. 
In that case, the relay can be selected for which $R_k$-$D$ link rate becomes maximum. 
The secrecy outage probability can be evaluated using law of total probability following 
the method of section \ref{subsec_traditional}  and \ref{subsec_subopt_re} as 
\begin{align}
\label{eq_23}
&P_o^{SS-RD}=\sum\limits_{k=1}^N \mathbb{P}\lb[\text{Relay}=R_k\rb]\mathbb{P}\lb[C_S^k<R_s\rb]\nn\\
&=\sum\limits_{k=1}^N\mathbb{P}\lb[\gamma_{kd}> \gamma_M^{-}\rb]
\mathbb{P}\lb[\gamma_{k}<\lambda\rb] \nn\\
&=\sum\limits_{k=1}^N\mathbb{P}\lb[\gamma_{kd}> \gamma_M^{-}\rb]
\lb(1-\mathbb{P}\lb[\min\lb(\gamma_{sk},\gamma_{kd}\rb)>\lambda\rb]\rb) \nn\\
&=\sum\limits_{k=1}^N\mathbb{P}\lb[\gamma_{kd}>\gamma_M^{-}\rb]
\lb(1-\lb(1-\mathbb{P}\lb[\gamma_{sk}<\lambda\rb]\rb)\rb.\nn\\ 
&\times\lb.\lb(1-\mathbb{P}\lb[\gamma_{kd}<\lambda\rb]\rb)\rb)\nn \\
&=\sum\limits_{k=1}^N\lb(\mathbb{P}\lb[\gamma_{kd}>\gamma_M^{-}\rb]\mathbb{P}\lb[\gamma_{sk}<\lambda\rb]
+\mathbb{P}\lb[\gamma_{kd}>\gamma_M^{-}\rb]\rb.\nn\\ 
&\times\lb.\mathbb{P}\lb[\gamma_{kd}<\lambda\rb]
-\mathbb{P}\lb[\gamma_{kd}>\gamma_M^{-}\rb]\mathbb{P}\lb[\gamma_{sk}<\lambda\rb]\mathbb{P}\lb[\gamma_{kd}<\lambda\rb]\rb) \\
\label{eq_23_A1}
&=\sum\limits_{k=1}^N\lb(\int_0^\infty\int_y^\infty f_{\gamma_{kd}}\lb(x\rb)f_{\gamma_M^{-}}\lb(y\rb)dxdy \rb. \nn\\
&\lb.\times\int_0^\infty\int_0^\lambda f_{\gamma_{sk}}\lb(x\rb)f_{\gamma_{ke}}\lb(y\rb)dxdy\rb. \nn \\
&\lb.+\int_0^\infty\int_0^\lambda\int_y^\lambda f_{\gamma_{kd}}\lb(x\rb)f_{\gamma_M^{-}}\lb(y\rb)f_{\gamma_{ke}}\lb(z\rb)dxdydz \rb. \nn \\
&\lb.-\int_0^\infty \lb( \int_0^\lambda f_{\gamma_{sk}}\lb(x\rb)dx
\times\int_0^\lambda\int_y^\lambda f_{\gamma_{kd}}\lb(x\rb) 
f_{\gamma_M^{-}}\lb(y\rb)dxdy \rb)\rb. \nn \\ 
&\times\lb.f_{\gamma_{ke}}\lb(z\rb)dz\rb),
\end{align}
where in this section 
\begin{align}
 \label{eq_23_1}
\gamma_M= \max_{\substack{i=1,\dots,N}}\{\gamma_{kd}\},~~ \text{and} ~~\gamma_M^{-}=\max_{\substack{i=1,\dots,N\\i\ne k}}\{ \gamma_{kd}\}. 
\end{align}
The distribution of $\gamma_M$ and $\gamma_M^{-}$ can be evaluated similarly following (\ref{eq_9})-(\ref{eq_14}) where 
$\beta_m^{\prime}=\sum_{l=1}^m\beta_{{i_l}d}$. 
After much simplification of  (\ref{eq_23_A1}), and solving the integrals, finally the outage probability expression can be shown in 
(\ref{eq_24}) where $\beta_m^{\prime}=\sum_{l=1}^m\beta_{{i_l}d}$.

This selection method is helpful when destination is selecting the relay. ICSI of all the relay-destination links can be obtained  
at the destination simply by channel estimation method \cite{book_proakis} from the ready-to-send (RTS) 
\cite{adam_adaptive_relay_selection} message sent from $R_k$ to $D$, for all $k=1, \cdots, N$.

\begin{figure*}[t!]
\hrule
\vspace{0.25 cm}
\centering The secrecy outage probability of traditional selection ($TS$) scheme obtained in \ref{subsec_traditional} 
with $\beta_m^{\prime}=\sum_{l=1}^m\beta_{i_l}$.

\begin{align}
\label{eq_16}
P_o^{TS}(R_s) &=\sum\limits_{k=1}^N \lb[-\frac{\alpha_{ke} e^{-(\rho-1)\beta_k}}{\beta_k\rho+\alpha_{ke}}
+ \sum\limits_{m=1}^{N-1} \lb( 
\frac{(-1)^m \mathbb {\sum}_m^{\prime} \beta_m^{\prime} \alpha_{ke}e^{{-\lb(\rho-1\rb)\lb(\beta_k+\beta_m^{\prime}\rb)}}}
{\lb(\beta_k+\beta_m^{\prime}\rb)\lb(\lb(\beta_k+\beta_m^{\prime}\rb)\rho
+\alpha_{ke}\rb)}   \rb.\rb.\nn\\
&\lb.\lb.-\frac{(-1)^m\mathbb {\sum}_m^{\prime}\beta_m^{\prime}}{\beta_k+\beta_m^{\prime}} 
- \frac{(-1)^m\mathbb {\sum}_m^{\prime}\alpha_{ke}e^{-\lb(\rho-1\rb)\lb(\beta_k+\beta_m^{\prime}\rb)}} 
{\lb(\beta_k+\beta_m^{\prime}\rb)\rho+\alpha_{ke}} \rb)\rb].
\end{align}
\hrule
\vspace{0.25 cm}
\centering The secrecy outage probability of suboptimal selection ($SS$-$RE$) scheme obtained in \ref{subsec_subopt_re} 
with $\beta_m^{\prime}=\sum_{l=1}^m\beta_{i_l}/\alpha_{le}$. 

\begin{align}
\label{eq_22}
P_o^{SS-RE}(R_s) &=\sum\limits_{k=1}^N \lb[-\frac{\alpha_{ke} e^{-(\rho-1)(\beta_k)}}{\beta_k\rho+\alpha_{ke}} 
+\sum\limits_{m=1}^{N-1} \lb( 
\frac{(-1)^m\mathbb {\sum}_m^{\prime}\beta_m^{\prime} \alpha_{ke}\lb[\frac{\beta_k}{\alpha_{ke}}+
\beta_m^{\prime}\rb]e^{-{\alpha_{ke}\lb(\rho-1\rb)}}}
{\lb(\frac{\beta_k}{\alpha_{ke}}+\beta_m^{\prime}\rb)\lb(\lb(\beta_k+\alpha_{ke}\beta_m^{\prime}\rb)\rho
+\alpha_{ke}\rb)}   \rb.\rb.\nn\\
&\lb.\lb.-\frac{(-1)^m\mathbb {\sum}_m^{\prime}\beta_m^{\prime}}{\frac{\beta_k}{\alpha_{ke}}+\beta_m^{\prime}} 
-\frac{(-1)^m\mathbb {\sum}_m^{\prime} \alpha_{ke}e^{-\lb(\rho-1\rb)\lb(\beta_k+\alpha_{ke}\beta_m^{\prime} \rb)}} 
{\lb(\beta_k+\alpha_{ke}\beta_m^{\prime}\rb)\rho+\alpha_{ke}} \rb)\rb].
\end{align}
\hrule
\vspace{0.25 cm}
\centering The secrecy outage probability of suboptimal selection ($SS$-$RD$) scheme obtained in \ref{subsec_subopt_rd} 
with $\beta_m^{\prime}=\sum_{l=1}^m\beta_{{i_l}d}$.

\begin{align}
\label{eq_24}
P_o^{SS-RD}(R_s)&=\sum\limits_{k=1}^N \lb[ -\frac{\alpha_{ke}e^{-(\rho-1)\beta_k}}{\beta_k\rho+\alpha_{ke}} 
+\sum\limits_{m=1}^{N-1}(-1)^m\mathbb {\sum}_m^{\prime} \lb(-\frac{\beta_m^{\prime}}{\beta_{kd}+\beta_m^{\prime}}
-\frac{\alpha_{ke}e^{-\lb(\rho-1\rb)\lb(\beta_k+\beta_m^{\prime} \rb)}}
{\rho\lb(\beta_k+\beta_m^{\prime} \rb)+\alpha_{ke}}\rb.\rb.\nn \\
&\lb.\lb.+\frac{\alpha_{ke}\beta_m^{\prime} e^{-\lb(\rho-1\rb)\lb(\beta_k+\beta_m^{\prime} \rb)}}
{\lb(\beta_{kd}+\beta_m^{\prime} \rb)\lb(\rho\lb(\beta_k+\beta_m^{\prime} \rb)+\alpha_{ke}\rb)}\rb)\rb].
\end{align}
\hrule
\end{figure*}

\subsection {Suboptimal Selection: Only S-R link ICSI is known ($SS$-$SR$)}
\label{sub_SS_SR}
To reduce the complexity and increase the network lifetime  of the system, as in section \ref{subsec_subopt_rd}, 
we introduce this partial relay selection method 
where only the source-relay link quality is used to select the best relay in secure communication setup. 
In a non secrecy communication setup, source-relay link quality based relay selection methods can be found  in 
\cite{krikidis_partial_relay_selection, rui_capacity_analysis_df}. 
When only the $S$-$R_k$, for all $k=1, \cdots, N$, link ICSI is known, relay can be 
selected only on the basis of source-relay rate as opposed to relay-destination rate. 
The secrecy outage probability can be obtained using law of total probability following similar method 
of the section \ref{subsec_subopt_rd}. 
Secrecy outage probability can be obtained as
\begin{align}
\label{eq_25}
&P_o^{SS-SR}=\sum\limits_{k=1}^N \mathbb{P}\lb[\text{Relay}=R_k\rb]\mathbb{P}\lb[C_S^k<R_s\rb]\nn\\
&=\sum\limits_{k=1}^N\mathbb{P}\lb[\gamma_{sk}> \gamma_M^{-}\rb]
\lb(1-\mathbb{P}\lb[\min\lb(\gamma_{sk},\gamma_{kd}\rb)>\lambda\rb]\rb) \nn\\
&=\sum\limits_{k=1}^N\lb(\mathbb{P}\lb[\gamma_{sk}>\gamma_M^{-}\rb]\mathbb{P}\lb[\gamma_{sk}<\lambda\rb]\rb. \nn\\
&\lb.+\mathbb{P}\lb[\gamma_{sk}>\gamma_M^{-}\rb]\mathbb{P}\lb[\gamma_{kd}<\lambda\rb]\rb.\nn \\
&\lb.-\mathbb{P}\lb[\gamma_{sk}>\gamma_M^{-}\rb]\mathbb{P}\lb[\gamma_{sk}<\lambda\rb]\mathbb{P}
\lb[\gamma_{kd}<\lambda\rb]\rb),
\end{align}
where in this section 
\begin{align}
 \label{eq_23_1} 
\gamma_M= \max_{\substack{i=1,\dots,N}}\{\gamma_{sk}\},
~\gamma_M^{-}=\max_{\substack{i=1,\dots,N\\i\ne k}}\{ \gamma_{sk}\},
\end{align}
and $\beta_m^{\prime}=\sum_{l=1}^m\beta_{s{i_l}} $.
The derivation procedure is similar to that of section \ref{subsec_subopt_rd}.
Even by looking at the similarity of (\ref{eq_23}) and (\ref{eq_25}), final outage probability can be obtained 
by simply replacing $\beta_{sk}$ with $\beta_{kd}$ and $\beta_{kd}$ with $\beta_{sk}$ in (\ref{eq_24}).
Final expression is not shown as it directly follows from  (\ref{eq_24}).

This selection method is a suboptimal one. The selection method is helpful when source is selecting a relay. 
ICSI of all the $S$-$R_k$, for all $k=1, \cdots, N$ can be 
obtained by clear-to-send (CTS) frame send from $R_k$ to $S$ \cite{adam_adaptive_relay_selection}.  The CTS method is also 
utilized in \cite{decentralized_pow_beaulieu}. 

\subsection {Proposed  Optimal Selection ($PS$)}
\label{subsec_proposed}
If no knowledge of ICSI is available except statistical information, this relay selection method is the optimal one. 
This selection method is proposed with the help of secrecy outage probability obtained for dual-hop single relay system in 
the section \ref{sec_out_prob}. The selection method is secrecy outage optimal as that relay is selected 
for which secrecy outage probability becomes minimum. Mathematically it can be expressed as 
\begin{align}
\label{eq_26}
k^*=\arg\min_{k\in[1,\cdots, N]}\lb(1-\frac{\alpha_{ke} 
e^{-\lb(\beta_{sk}+\beta_{kd}\rb)\lb(\rho-1\rb)}}{\rho \lb(\beta_{sk}+\beta_{kd}\right)+\alpha_{ke}}\rb).
\end{align}
Measuring the secrecy outage probabilities, $P_o^k(R_s)$, of all the individual single relay systems as obtained in 
(\ref{eq_6}), optimum relay $k^*$ can be found. This selection method only requires 
all the links' statistics for the outage measurement. This relay selection method does not require any link ICSI, 
so complex channel measurements are not necessary hence reduces power consumption. 
This is a one-time process as channel statistics does not considerably 
change over time compared to the ICSI of links. This relay selection method can improve the 
secrecy even when there is severe constraint on resource like power and computational complexity.
\section {Asymptotic Analysis}
\label{sec_asymp}
The behaviour of secrecy outage probability is important for system design 
when source-relay and or relay-destination link SNRs are increased asymptotically as compared to eavesdropper's link.
This can happen if source-relay and or relay-destination are very closely placed as compared to the eavesdropper. The 
received SNR can also increase at the relay or the destination due to increased transmit power or better fading channel between 
them as compared to that of eavesdropper's. Asymptotic analysis provides simpler expression
as a function of constituent parameters to understand the behaviour at a limiting case of high SNR
with the variation of those parameters. In this section, asymptotic analysis 
is provided for secrecy outage probability of 
single relay system obtained in the section \ref{sec_out_prob} and multiple relay systems with relay selection.
Two cases are of main importance, 1) when $S$-$R_k$ and $R_k$-$D$ link average SNRs are same,  for all $k$, 
and together tends to infinity, i.e. $1/\beta_{sk}=1/\beta_{kd}=1/\beta\rightarrow  \infty$, we call it as
balanced case, 2) when either of the $S$-$R_k$ or $R_k$-$D$ for all $k$, link average SNR tends to infinity we call it as unbalanced case, 
 i.e. $1/\beta_{sk}$ is fixed and $1/\beta_{kd}=1/\beta\rightarrow  \infty$, or 
$1/\beta_{kd}$ is fixed and $1/\beta_{sk}=1/\beta\rightarrow  \infty$.  
\subsection{Single Relay: Balanced Case}
\label{asymp_single_balance}
For the balanced case when $1/\beta_{sk}=1/\beta_{kd}=1/\beta\rightarrow  \infty$, the secrecy outage probability of dual-hop single relay system 
in (\ref{eq_6}) can be expressed as 
\begin{align}
\label{eq_27}
P_o^{k(AS)}(R_s)
=&\frac{2}{\frac{1}{\beta}}\lb[\frac{\rho}{\alpha_{ke}}+\lb(\rho-1\rb)\rb] .
\end{align}
This shows that at a very high main channel SNR ($1/\beta$), outage probability is inversely proportional to the main channel 
SNR and it tends to zero. It is directly proportional to the eavesdropper channel SNR ($1/\alpha_{ke}$) and required 
threshold secrecy rate $\rho$.

Diversity order is an important measure of how fast the secrecy outage is decreasing as SNR tends to infinity. 
It provides an intuitive understanding into the impact of the number of relays on the secrecy outage probability.
The standard definition of the diversity order \cite{Zou_Wang_Shen_optimal_relay_sel, DBLP_Zou_Outage_Cognitive_Radio} is 
\begin{align}
\label{eq_27_1}
d=  -\underset{\text{SNR} \rightarrow \infty}{\text{lim}}  \frac{\log P_o(\text{SNR}) }{\log (\text{SNR})},
\end{align}
where $P_o(\text{SNR})$ is the outage probability as a function of $\text{SNR}=1/\beta$. Using this definition, diversity order of 
(\ref{eq_27}) can be obtained as one. Diversity order, $d$, is also same as the power of the SNR at the denominator of 
(\ref{eq_27}) (or the slope of the curve in log graph).
The system achieves diversity order of one, which is also intuitive as there is no relay selection. 

Now, we will find how the relative performance depends with eavesdropper channel quality improvement 
and $\rho$. We will find relatively how much main channel SNR, $1/\beta$, is required in decibel 
(dB) to achieve a given $P_o(R_s)$,  when eavesdropper channel average channel SNR improves 
from $1/\alpha_1$ to $1/\alpha_2$ at different rates of $\rho_1$ and $\rho_2$  
respectively. Here $1/\alpha_1<1/\alpha_2$ and $\rho_1<\rho_2$. 
Here we have dropped the subscript of $\alpha$ and $\beta$, and have used new subscript `$1$' and `$2$' to identify 
first and second realization of same $\alpha$ and $\beta$. Let us find the main channel SNR required in dB from 
(\ref{eq_27}) at $\rho_1$ and $\rho_2$ respectively as
\begin{align}
\label{eq_27_A1}
G_1&=\frac{1}{\beta_2}\Big{|}_\text{dB} -\frac{1}{\beta_1}\Big{|}_\text{dB}= 
10\log_{10}\lb[ \frac{\frac{1}{\alpha_2}+\frac{\rho_1-1}{\rho_1}}{\frac{1}{\alpha_1}+\frac{\rho_1-1}{\rho_1}}\rb],\nn\\
G_2&=\frac{1}{\beta_2}\Big{|}_\text{dB} -\frac{1}{\beta_1}\Big{|}_\text{dB}= 
10\log_{10}\lb[ \frac{\frac{1}{\alpha_2}+\frac{\rho_2-1}{\rho_2}}{\frac{1}{\alpha_1}+\frac{\rho_2-1}{\rho_2}}\rb].
\end{align}
Where $1/\beta_1|_\text{dB}$ is the dB equivalent of $1/\beta_1$. If $G_1>G_2$, following must be true
\begin{align}
\label{eq_27_A2}
\frac{\frac{1}{\alpha_2}+\frac{\rho_1-1}{\rho_1}}{\frac{1}{\alpha_1}+\frac{\rho_1-1}{\rho_1}}&>
\frac{\frac{1}{\alpha_2}+\frac{\rho_2-1}{\rho_2}}{\frac{1}{\alpha_1}+\frac{\rho_2-1}{\rho_2}}\nn\\
\Rightarrow \frac{\frac{1}{\alpha_2}-\frac{1}{\alpha_1}}{\frac{1}{\alpha_2}+1- \frac{1}{\rho_1}} &>
\frac{\frac{1}{\alpha_2}-\frac{1}{\alpha_1}}{\frac{1}{\alpha_2}+1-\frac{1}{\rho_2}}.
\end{align}
We can see from (\ref{eq_27_A2}) that as $\rho_2>\rho_1$, (\ref{eq_27_A2}) is true, hence $G_1>G_2$. This analysis says 
that improvement in eavesdropper channel quality degrades the secrecy outage more when required threshold 
secrecy rate is low than rate is high. 
Conversely, relatively higher main channel SNR is required to compensate for the 
improvement in eavesdropper link quality at lower rate requirement than at higher rate requirement. 
\subsection{Single Relay: Unbalanced Case}
\label{subsec_unbal}
For the unbalanced case, we study the behavior of secrecy outage probability in (\ref{eq_6}) keeping the average 
SNR of the source-relay link fixed and  
asymptotically increasing the average SNR of the relay-destination link. When
$1/\beta_{sk}$ is fixed and $1/\beta_{kd}=1/\beta\rightarrow  \infty$,  
the asymptotic secrecy outage probability can be expressed as a summation of a constant quantity and an asymptotically 
varying term with $1/\beta$ as 
\begin{align}
\label{eq_28}
P_o^{k(AS)}(R_s)&= \lb[1-\frac{\alpha_{ke}e^{-\beta_{sk}\lb(\rho-1\rb)}}{\rho\beta_{sk}+\alpha_{ke}} \rb] \nn\\
&+\frac{1}{\frac{1}{\beta}}\lb[\frac{\rho+(\rho-1)\alpha_{ke}e^{-\beta_{sk}(\rho-1)}}{\lb(\rho\beta_{sk}+\alpha_{ke}\rb)}\rb].
\end{align}
At low SNR asymptotically varying term dominates but vanishes at high SNR. 
When $1/\beta_{kd}$ is fixed and $1/\beta_{sk}=1/\beta\rightarrow  \infty$,  the asymptotic secrecy outage 
probability can be expressed same as in (\ref{eq_28}) with $\beta_{sk}$ replaced with $\beta_{kd}$. 
This says that the system is symmetric for both the unbalanced cases whether 
$1/\beta_{sk}$ is fixed or $1/\beta_{kd}$ is fixed. From (\ref{eq_28}) it can be understood that 
unbalance is caused due to fixing average SNR of any hop  in dual-hop system. This limits the secrecy outage probability to a constant 
even if average SNR of the other hop is infinitely increased. 
\subsection{Optimal Selection: Balanced Case}
\label{subsec_optim_bal}
In the balanced case, asymptotic expression of secrecy outage probability in (\ref{eq_7}) 
for the optimal relay selection,  can be evaluated  as
\begin{align}
\label{eq_29}
P_o^{OS(AS)}(R_s)= \frac{2^N}{\frac{1}{\beta^N} }\prod_{k=1}^N\lb[\frac{\rho}{\alpha_{ke}}+\lb(\rho-1\rb)\rb].
\end{align}
Comparing (\ref{eq_27}) with (\ref{eq_29}) we can see that asymptotic expression for secrecy outage probability of optimal 
relay selection is the product of individual asymptotic expressions of single relay system.
It can be seen that the denominator contains power of $N$ at main channel  $\text{SNR} = 1/\beta$ . 
From  (\ref{eq_27_1}) and (\ref{eq_29}), diversity order can be obtained as $d=N$. 
We can conclude that the diversity order of $N$ can be achieved when a single relay is chosen from a 
set of $N$ relays which is also intuitive.
\subsection{Optimal Selection: Unbalanced Case}
\label{subsec_optim_unbal}
In the unbalanced case of $1/\beta_{sk}$ is fixed and 
$1/\beta_{kd}=1/\beta\rightarrow  \infty$,  for all $k=1, \cdots, N$, 
the secrecy outage probability of optimal relay selection  in (\ref{eq_7}) tends to a constant value as given by
\begin{align}
\label{eq_30}
P_o^{OS(AS)}(R_s)=\prod_{k=1}^N\lb(1-\frac{\alpha_{ke}e^{-\beta_{sk}\lb(\rho-1\rb)}}{\rho\beta_{sk}+\alpha_{ke}}\rb).
\end{align}
Here we have not shown the asymptotic varying term which can also be obtained as in (\ref{eq_28}).
When $1/\beta_{kd}$ is fixed and 
$1/\beta_{sk}=1/\beta\rightarrow  \infty$,  for all $k=1, \cdots, N$, the constant value can be 
obtained by replacing  $\beta_{sk}$ with $\beta_{kd}$ in (\ref{eq_30}) for all $k=1, \cdots, N$. 
Comparing (\ref{eq_28}) and (\ref{eq_30}) we can observe that the constant value of secrecy outage probability for 
optimal relay selection is the product of constant values of individual single relay systems. As each constant 
value in (\ref{eq_28}) is less than unity, the optimal relay selection always improves the performance. 

By following section \ref{subsec_unbal} and section \ref{subsec_optim_unbal} it can be concluded that either 
of the source-relay or relay-destination link quality can equally affect the secrecy outage performance. 
Literature in \cite{krikidis_iet_opport_rel_sel, krikidis_twc_Rel_Sel_Jam, 
Bao_Relay_Selection_Schemes_Dual_Hop_Security, Alotaibi_Relay_Selection_MultiDestination, 
Poor_Security_Enhancement_Cooperative, Hui_Secure_Relay_Jammer_Selection} 
neglect the effect of source-relay link quality by considering high SNR scenario for 
the first hop or perfect decoding at the DF relays. The effects of both the source-relay or relay-destination 
link quality can be understood by consideration both of them for the derivation of the performance which is 
done in this paper.
\subsection{Traditional Selection: Balanced Case}
Derivation of the asymptotic secrecy outage probability for the balanced case of 
traditional selection in (\ref{eq_16}) is not straight forward as in section 
\ref{asymp_single_balance} and \ref{subsec_optim_bal} due to the complexity of the PDF of $\gamma_M^{-}$. 
For this we have first found the asymptotic PDFs and CDFs required to derive (\ref{eq_12}).
As average SNR is very high, i.e. $1/\beta_k \rightarrow \infty$, we use the approximation 
$\exp(-\beta_k x)\approx (1-\beta_k x)$ as $\beta_k\rightarrow 0$ for the derivation of 
asymptotic PDFs and CDFs. We use 
\begin{align}
\label{eq_30_1}
F_{\gamma_k}(x)\approx \beta_k x, 
\end{align}
\begin{align}
\label{eq_30_2}
F_{\gamma_M^{-}}(x)\approx \prod_{\substack{i=1 \\
 i\ne k}}^N\beta_i x^{N-1},  
\end{align}
in (\ref{eq_12}). The asymptotic secrecy outage probability, $P_o^{TS(AS)}(R_s)$, of the traditional 
relay selection method can then be derived 
for the balanced case when $1/\beta_k=1/\beta$, for all $k=1,\cdots, N$, as
\begin{align}
\label{eq_30_5}
&P_o^{TS(AS)}(R_s) \nn\\&=\frac{1}{\frac{1}{\beta^N}}\sum\limits_{k=1}^N\lb[\sum\limits_{i=1}^N 
\sum\limits_{j=0}^N \binom{N}{j}\frac{(\rho-1)^j \rho^{N-j} \Gamma(N-j+1)}{N\alpha_k^{N-j}}\rb].
\end{align}
Applying (\ref{eq_27_1}),  diversity order of $P_o^{TS(AS)}(R_s)$ in (\ref{eq_30_5}) 
can be evaluated as $d=N$.

\section{Numerical Results} 
\label{sec_results}

This section describes the analytical results along with the simulation. 
From all the figures it can be seen that analytical results exactly match the simulated ones. 
The unit of threshold secrecy rate $R_s$ is assumed to be bits per channel use (bpcu). 
Noise power at all the nodes are assumed to be same. Low and high required rate of $R_s=0.1$ and $R_s=2.0$ are assumed 
to cover reasonable range of threshold secrecy rate.

\begin{figure}
\centering
\includegraphics[width=0.48\textwidth] {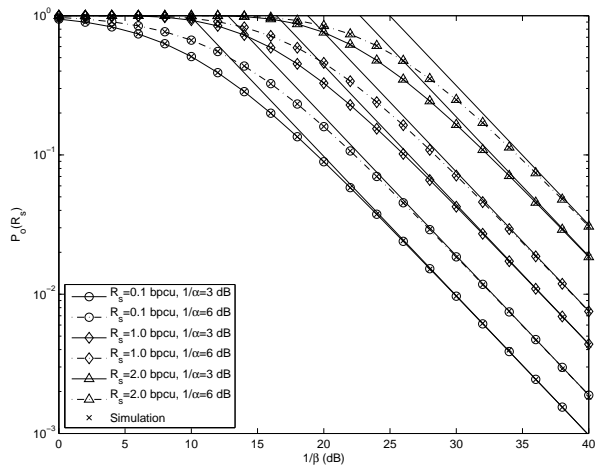}
\vspace*{-0.4cm}
\caption{Secrecy outage probability of single relay balanced case.}
\label{FIG_2}
\vspace{-0.3cm}
\end{figure}

Fig. \ref{FIG_2} shows the secrecy outage probability, $P_o(R_s)$, of the dual-hop single relay system expressed in 
(\ref{eq_6}) for the balanced case with total SNR $1/\beta$. Total power is divided equally among the $S$ and $R_k$ i.e. individual SNRs
at $S$ and $R_k$ becomes $1/\beta_{sk}=1/\beta_{ke}=1/(2\beta)$ considering equal noise power at each terminals. 
The figure is plotted with different $R_s=0.1, 1.0, 2.0$  and relay to eavesdropper average 
SNR $1/\alpha_{ke}=1/\alpha=3, 6$ dB. 
Corresponding asymptotic analysis expressed in (\ref{eq_27}) is also shown by solid straight lines passing through the curves.  
It is observed from the figure that spacing between asymptotic straight lines for $1/\alpha=3$ dB and $1/\alpha=6$ dB, at 
a given $P_o(R_s)$, is more for low $R_s=0.1 $  and subsequently decreases for $R_s=1.0$ and $R_s=2.0$ . 
This confirms the analysis shown in (\ref{eq_27_A2}) which says that improvement in eavesdropper channel quality degrades 
the $P_o(R_s)$ more when required $R_s$ is less.

\begin{figure}
\centering
\includegraphics[width=0.48\textwidth] {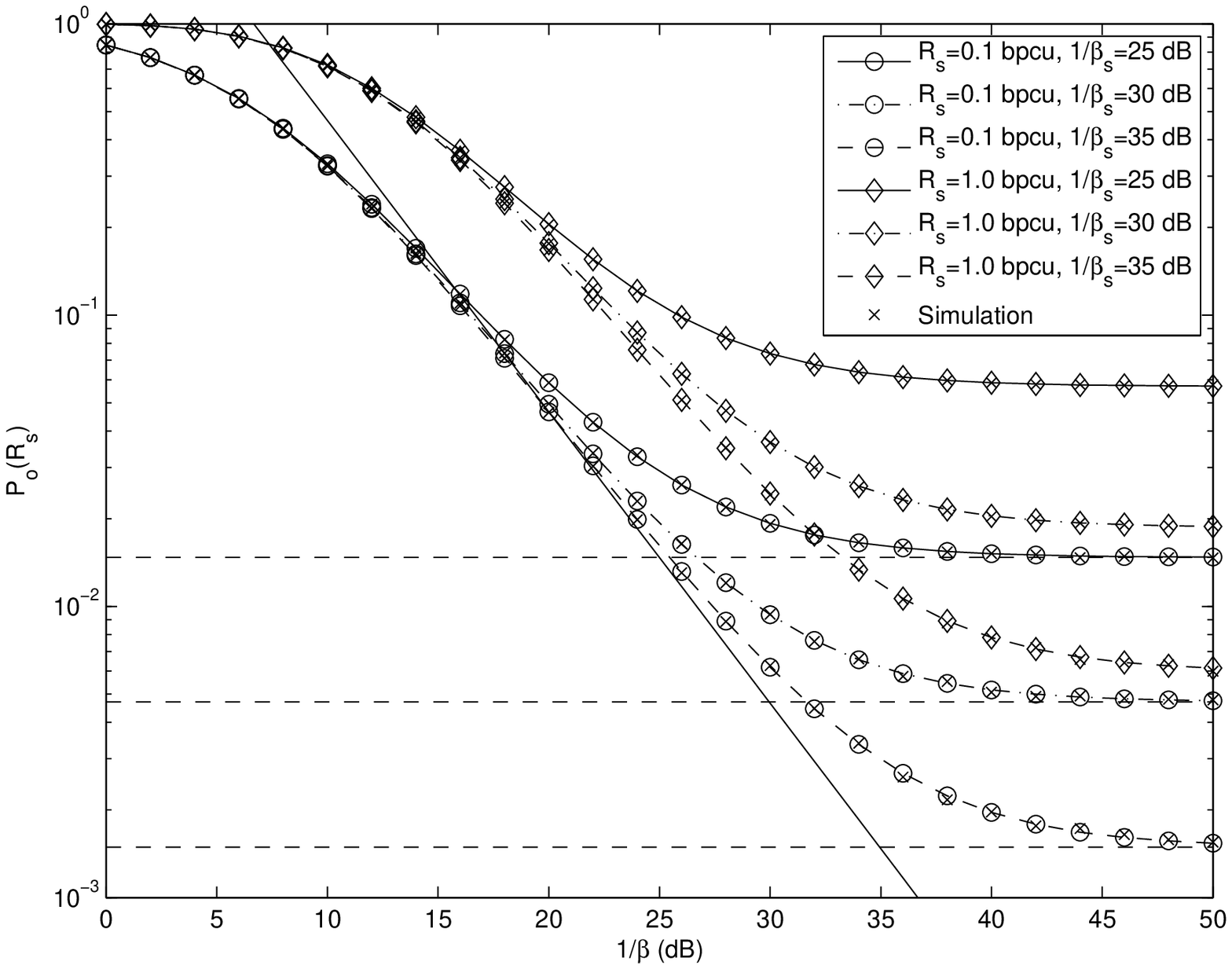}
\vspace*{-0.4cm}
\caption{Secrecy outage probability of single relay unbalanced case.}
\label{FIG_3}
\vspace{-0.3cm}
\end{figure}

In Fig. \ref{FIG_3}, $P_o(R_s)$ is plotted for the dual-hop single relay system expressed in (\ref{eq_6}) for the 
unbalanced case with average SNR of $1/\beta_{kd}=1/\beta$ at a given $1/\beta_{sk}=1/\beta_s= 25, 30, 35$ dB 
and $1/\alpha_{ke}=1/\alpha= 6$ dB. 
It is observed that $P_o(R_s)$ tends to a fixed constant 
for a given $1/\beta_s$, even if $1/\beta$ increases. The fixed constants are shown only for the $R_s=0.1$  
with horizontal dashed line, which is derived in (\ref{eq_28}). It is also observed but not shown in the figure 
that by keeping $1/\beta_{kd}=1/\beta_d$ fixed and by increasing $1/\beta_{sk}=1/\beta$ 
produces the same result. This is also discussed in the section \ref{subsec_unbal}. 
This flooring of curves means that the secrecy outage 
probability is constrained by either of the $S$-$R_k$ or $R_k$-$D$ link quality by identical manner. 
It is interesting to observe that the asymptotically varying term of (\ref{eq_28}) shown as straight solid line,
 crosses dashed lines at $1/\beta= 25, 30, 35$ dB which is exactly the same fixed $1/\beta_s= 25, 30, 35$ 
dB for which the figures are drawn. This is precisely the point after which average SNR of the hop exceeds the 
average SNR of the other hop.

\begin{figure}
\centering
\includegraphics[width=0.48\textwidth] {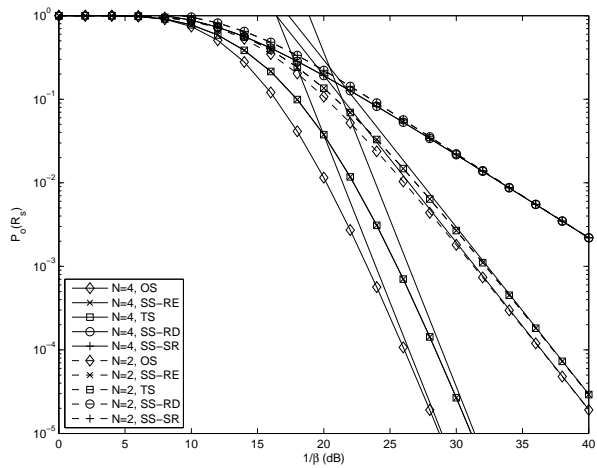}
\vspace*{-0.4cm}
\caption{Secrecy outage probability comparison of various relay selection schemes with $N$.}
\label{FIG_4}
\vspace{-0.3cm}
\end{figure}

In Fig. \ref{FIG_4}, comparison of $P_o(R_s)$ for various relay selection schemes are presented  with different number of relays 
$N=2$ and $N=4$ in the system.  The comparison is shown for $R_s=1.0$  when $1/\beta_{sk}=1/\beta_{ke}=1/(2\beta)$ and 
$1/\alpha_{ke}=1/\alpha=3$ dB for all $k$. The asymptotic analysis of $OS$ and $TS$ schemes for balanced case  
in (\ref{eq_29}) and (\ref{eq_30_5}) respectively are plotted with solid straight line through the curved ones for $N=2, 4$. 
Clearly increase in $N$ improves the secrecy outage probability 
as diversity increases i.e. slope of the solid straight lines increase from $N = 2$ to $N = 4$. 
For example  balanced case of $OS$ and  $TS$  both scheme achieves the diversity order of 
$N$, when $N$ relays are used as can be seen from (\ref{eq_29}) and (\ref{eq_30_5}). 
It is seen that the gap between asymptotic straight lines for $OS$ and  $TS$ are more for $N = 4$ than $N = 2$. 
This suggests that improvement is more by applying the optimal relay selection as compared to the traditional relay selection 
 when the selection is carried out for larger number of available relays.
It is observed that the performances of $SS$-$RE$ and $TS$ schemes are same. This is due to the fact that all the eavesdropper 
links have identical average SNR. 
So knowledge of average SNR of $R_k$-$E$ link can not give any advantage over traditional relay selection. 
Following the same reason, the performance that of $SS$-$RD$ and $SS$-$SR$ schemes are also same as 
$S$-$R_k$ link and $R_k$-$D$ link average SNRs are same.

\begin{figure}
\centering
\includegraphics[width=0.48\textwidth] {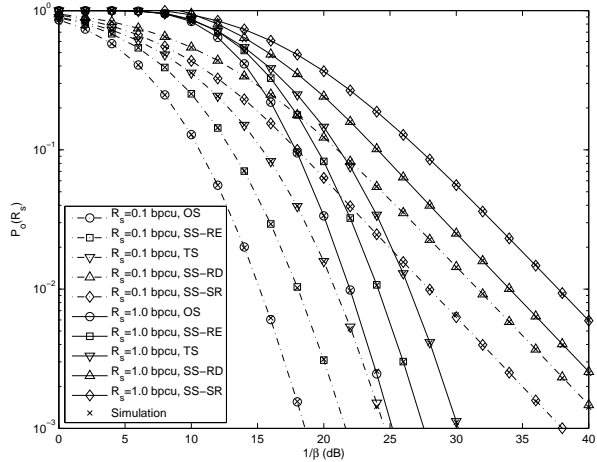}
\vspace*{-0.4cm}
\caption{Secrecy outage probability comparison of various relay selection schemes with different eavesdropper link SNRs.}
\label{FIG_5}
\vspace{-0.3cm}
\end{figure}

In Fig. \ref{FIG_5}, comparison of $P_o(R_s)$ for various relay selection schemes are given when $N=4$ and $1/\alpha_{ke}$ are different
for all $k$, i.e. $1/\alpha_{ke}= 0, 3, 6, 9$ dB for $k=1, 2, 3, 4$ respectively, at $R_s = 0.1$ and $R_s = 1.0$. 
At $R_s = 0.1$, curves are plotted 
when $30\%$ of total power $1/\beta$ is allotted to $S$-$R_k$ links for all $k$  and $70\%$ to $R_k$-$D$ for all $k$, 
whereas at $R_s = 1.0$, when $70\%$ of $1/\beta$ is assigned to $S$-$R_k$ links for all $k$  and 
$30\%$ to $R_k$-$D$ for all $k$. It is observed that the performance of 
$OS$ scheme is the best and the order of performances from the best to the 
worst can be identified as $OS$, $SS$-$RE$, $TS$, $SS$-$RD$ or $SS$-$SR$. 
$SS$-$RE$ works better than $TS$ as it is the improved version of $TS$. Improvement is achieved by utilizing the knowledge  
of average SNR of $R_k-E$ links. Depending on the percentages 
of power allocated to the $S$-$R_k$ or $R_k$-$D$ links,  any one of $SS$-$RD$ or $SS$-$SR$ scheme becomes worst. 
When less power is allocated to $S$-$R_k$  than $R_k$-$D$ links, i.e. for $R_s=0.1$ in figure, $SS$-$SR$ performs better than $SS$-$RD$.
When less power is allocated to $R_k$-$D$ than $S$-$R_k$ links, i.e. for $R_s=1.0$ in figure, $SS$-$RD$ performs better than $SS$-$SR$.
This can be explained by noting that in a dual-hop DF relay system, the capacity is limited by the worst channel between two hops, 
in other words by the hop having least average SNR as can be seen from (\ref{eq_4}). Equation (\ref{eq_4}) is used in (\ref{eq_5}). 
That is why selection among worse channels performs better than selection among better channels.

\section{Conclusions}  
\label{sec_conclusion}
In this paper,  secrecy outage probability of a dual-hop regenerative single relay system is obtained.  
Secrecy outage probabilities of different relay selection schemes in a dual-hop regenerative multi-relay 
system, which require ICSI, are derived. Partial relay selection schemes based only on source-relay 
or relay-destination ICSI are introduced  to reduce the power consumption and increase the lifetime of the network. 
Corresponding secrecy outage is also evaluated. 
An optimal relay selection scheme, which does not require any ICSI, is proposed based on the 
secrecy outage probability. Secrecy outage probabilities in all the cases are obtained in closed-form 
without assuming high SNR scenario. Asymptotic and diversity order analysis of secrecy outage probability is presented 
for single relay and multi-relay system with relay selection when there is balance and unbalance in the dual-hop link. 
We observe that improvement in eavesdropper channel 
quality has greater effect on secrecy outage at lower required rate than at higher 
required rate. We also observe that optimal relay selection improves the performance as compared to the 
traditional relay selection more when number of relays are more. 
It is interesting to find that either of the source-relay or relay-destination 
link quality can equally limit the secrecy outage performance.

\bibliographystyle{IEEEtran}
\bibliography{IEEEabrv,MYALL_REFERENCE}
\end{document}